\def \bfr {\begin{flushright}}
\def \efr {\end{flushright}}
\def \caja {\makebox[3.2cm][1]}
\def \d {\hbox{d}\,}
\def \square {\hbox{$\sqcup\!\!\!\!\sqcap$}}
\def \s {\hbox{s}}

\def \v {\vskip}

\def \p {\partial}

\def \ba {\begin{array}}
\def \ea {\end{array}}

\def \bea {\begin{eqnarray}}
\def \eea {\end{eqnarray}}

\def \be {\begin{equation}}
\def \ee {\end{equation}}

\def \bfr {\begin{flushright}}
\def \efr {\end{flushright}}
\def \caja {\makebox[3.2cm][1]}
\def \E {{\bf E}}
\def \B {{\bf B}}
\def \H {{\bf H}}
\def \D {{\bf D}}
\def \rot {\hbox{rot}}
\def \div {\hbox{div}}
\documentstyle[12pt,a4]{article}
\begin{document}

%%%%%%%%%%%%%%%%%%%%%%%%%%%%%%%%%%%%%%%%%%%%%%%%%%%%%%%%%%%%%%
\pagestyle{empty}
\bfr

\caja{{\bf Imperial-TP/93-94/30}}

\caja{March 1994}
\efr
\v 1cm
\begin{center}

{\bf OPTICS,  MECHANICS AND QUANTIZATION
OF REPARAMETRIZATION INVARIANT SYSTEMS}
\footnote[2]{
Work partially supported by the D.G.I.C.Y.T.}
\v 0.3cm

Miguel Navarro$^{1,2,3}$, Julio Guerrero$^{2,4}$, \\
and V\'\i ctor Aldaya$^{2,3}$
\v 0.3cm
\today
\end{center}

\noindent 1.- The Blackett Laboratory, Imperial College, Prince
Consort Road, London SW7 2BZ;
United Kingdom.

\noindent 2.- Instituto Carlos I de F\'\i sica Te\'orica y Computacional,
Facultad  de  Ciencias, Universidad de Granada, Campus de Fuentenueva,
18002. Granada.

\noindent 3.- IFIC, Centro Mixto Universidad de
Valencia-CSIC, Burjassot 46100-Valencia, Spain.

\noindent 4.- Departamento  de  F\'\i sica  Te\'orica  y  del  Cosmos,
Facultad  de  Ciencias, Universidad de Granada, Campus de Fuentenueva,
18002. Granada.

\centerline{\bf Abstract}

In this paper we regard the dynamics obtained from the Fermat
principle as being the
classical theory of light. We (first-)quantize the action and
show how close we can get to the Maxwell theory. We show that
quantum Geometric Optics is not a theory of fields in curved
space.
Considering Classical Mechanics to be on the same footing, we
show the parallelism between Quantum Mechanics and Quantum
Geometric Optics. We show that, due to the reparametrization
invariance of the classical theories, the dynamics of the quantum
theories is given by a Hamiltonian constraint.
 Some implications of the above analogy in
the quantization of true reparametrization-invariant theories
are discussed.

\vfil\eject

%%%%%%%%%%%%%%%%%%%%%%%%%%%%%%%%%%%%%%%%%%%%%%%%%%%%%%%%%%%%%%%%%%%

\setcounter{page}{1}
\pagestyle{plain}

\section{Introduction.}

The theory that can be properly identified with a
classical theory of Optics  is Geometric  Optics,  which is
obtained from the requirement for light paths to be of
extremal  optical  length  \cite{[Born]}.
This theory was used to  describe successfully the dynamics  of
light until Electrodynamics were formulated and the relationship
between  this theory and light discovered.  Since then, Optics
has  been  treated  as  a  chapter  of  Electrodynamics  and  its
quantization  has been achieved as  a  byproduct  of
Quantum  Electrodynamics.  Hence, the history of Optics
has been  quite different from that of Mechanics, despite the close
points  of  departure (a classical action in both cases)  and  the
common  point of arrival (Quantum Electrodynamics in both  cases). This
paper is meant to be a small step in filling this gap.
In fact, the basics of this paper are similar to the ones that
motivated the introduction of Quantum Mechanics on the
grounds of its
analogy to wave optics. Nevertheless,
to our knowledge, no development similar to the present one
has been made before.

In this paper we study firstly (Sec. 3) the theory
obtained by
(canonically) quantizing Geometric Optics as given by the
Fermat principle\footnote[2]
{In some literature on the topic,
the procedure that we call ``quantization" is referred to as
``wavization"  [see for instance
Ref. \cite{[Wolf]} and references therein].}.

We shall consider two distinct interpretations of Geometric Optics:
it describes a  particle with constant mass moving in a
{\it curved} Euclidean $3+0$-dimensional space(-time) or a particle
moving in {\it flat} Euclidean $3+0$-dimensional space(-time) but
with a site-dependent mass.  We show that, although
both interpretations can be given to the
same classical dynamics, they lead to different quantum theories.
In fact,  quantizing the classical theory
{\it \`a la} Proca  we
find that the good interpretation, i.e. the one that
approximates Maxwell theory, is the theory of a particle with
site-dependent mass. Hence we find that,
in contradiction with na\"\i ve expectations,  Quantum Geometric Optics
cannot be identified with a theory of fields in curved space.

The above discussion leads us naturally to discuss the case of
mechanical systems. In Sec. 4 we discuss the optical analogy of
Classical Mechanics. We show that a complete parallelism can be
established between Geometric Optics and ``time-independent"
Classical Mechanics, an analogy which can be carried throughout
the quantization procedure, to the quantum
theory. We show that
Geometric Optics and the optical ``image" of Classical Mechanics
are described by reparametrization-invariant systems.
Hence their quantum dynamics
are tied to a Hamiltonian constraint. As a consequence,
Classical Mechanics or Geometric Optics describes systems with
Hamiltonian constraints whose good order is known, as well
as the physical interpretation of the associated quantum
theories.
In Sec. 5 we discuss some implications of these results in the
quantization of
true reparametrization-invariant systems, such as gravity
theories.

Before pursuing the central aims of this paper we
will describe briefly, in Sec. 2, the classical dynamics of
(Geometric) Optics.

\section{The classical dynamics of Optics.}

The classical action of Optics,  i.e.,  the action which gives  the
dynamics of the so-called Geometric Optics, is given by
\cite{[Born]}:

\be  S = \int n(x) \d s =
\int n\sqrt{{\bf{\dot x}}^2}\d  \tau\label
{1}\ee
where $\tau$ parametrizes the trajectory.
In this action $n$, the refraction index, is physically identified
with  the quotient  $c/v$ where $c$ $(v)$ is the speed of  light  in  the
vacuum (medium).  However, unlike  the  mechanical
case, $v$ ($n$) is not a degree of freedom but a datum which must be
given in advance. Since
$v$ is the only reference to the physical time which
appears in the action (\ref{1}), it can be said that the action
(\ref{1})  corresponds  to a frozen  theory,  i.e.,  a  theory
without time.
The  action   (\ref{1})   is    invariant    under
reparametrization: $\tau\rightarrow \tau(\xi)$, and thus,
the  study of its classical dynamics can  proceed  in
two  differents ways:  a) A non-reparametrization-invariant  study  which
begins  by  fixing  the parametrization of the trajectory and  b)
a reparametrization-invariant, or manifestly covariant,
approach  which
does not require such a choice.

\subsection{Non-explicitly covariant approach.}
Let us choose as ``time", or preferred direction of motion,
the  co-ordinate z.
The action (\ref{1}) takes the form

\be S   =\int L \d z
 =   \int   \d   z
n(x,y,z)\sqrt{1  +   {\dot   x}^2   +
{\dot y}^2}\label{Sdosprima}\ee

The phase space is given by the co-ordinates $x, y$ and the
momenta

\be P_x = \frac{\partial L}{\partial \dot x} =
\frac{n^2{\dot x}}{n\sqrt{1  +   {\dot   x}^2   +
{\dot y}^2}}\ee
\be P_y = \frac{\partial L}{\partial \dot y} =
\frac{n^2{\dot y}}{n\sqrt{1  +   {\dot   x}^2   +
{\dot y}^2}}\ee

The Hamiltonian ${\cal H}$ takes the form

\be {\cal H} =  P_x \dot x + P_y \dot y - L = -\sqrt{n^2(x,y,z) -
P^2_x - P^2_y}\label{H}\ee

The  equations of motion for any function $F$ on the phase  space  are
given, as usual, by

\be \frac{\d F}{\d z} = \{F, {\cal H}\}\>, \ee
where  the  Poisson bracket is determined by the  symplectic  form
$\omega$,
which in the coordinates defined above
has a Darboux form:

\be \omega = \d P_x\wedge\d x +  \d P_y\wedge\d y\>.\ee

\subsection{Manifestly covariant description.}

The action (\ref{1}) has  dimensions of length.  In order  to
get  an  action with the adequate dimensions, it is convenient to
multiply it by a factor $\hbar/\lambda$ and take

\be  S = \frac\hbar\lambda\int n(x) \d s =
\frac\hbar\lambda\int n\sqrt{{\bf{\dot x}}^2}\d  \tau
\label {8}\ee
Here $\hbar$  is  Plank's
constant  and $\lambda$ should be identified with the wavelength
of light.
The particular form of this factor clearly
indicates  that the  theory
we are studying describes only light with a definite wavelength
or frequency. Hence, in the present theory, photons with different
wavelengths will behave as different particles and will not interact at
all  with  photons  with a different  wavelength.  For  the  same
reason,
the interaction with the medium, as described by the present
theory, will preserve the wavelength of light.

The action (\ref{8}) essentially admits two different
interpretations, which we will refer to in the sequel as first and
second interpretations:

\noindent 1) It can be interpreted as a theory of a
particle with site-dependent mass $\frac\hbar\lambda n$
evolving in a flat Euclidean $3+0$-dimensional space(-time).

\noindent 2) If we introduce in (\ref{8})
the refraction index into the square root we get the action of a
particle  with  mass $\hbar/\lambda$ moving in  an  euclidean  $3+0$
dimensional space(-time) with a conformally flat metric

\be \d \s^2 = n^2 (\d x^2 + \d y^2 + \d z^2)\>.\label{9}\ee
This analogy, together with the fact that in Electrodynamics in
material media a tensor appears,
the dielectric tensor $\epsilon_{ij}$ having the form
 $\epsilon_{ij}= n^2\delta_{ij}$ for isotropic media,
leads us  to generalize and interpret Geometric Optics
as  the  dynamics of a particle with constant mass $\frac\hbar\lambda$
moving in  a  curved (Euclidean)  $3+0$
dimensional space(-time).

We shall study  both interpretations at the same time
by considering the action of a particle with
site-dependent mass $\frac\hbar\lambda m$
moving in a curved $3+0$-dimensional space(-time)
with metric $N_{ij}$. Its action is

\be S = \frac{\hbar}{\lambda} \int m\d s
=\frac{\hbar}\lambda \int \d\tau  m\sqrt{N_{ij}\dot x^i\dot x^j}
\label{12}\ee

The interpretation in 1) is recovered by putting $m=n\>,
N_{ij}=\delta_{ij}$. The interpretation in 2)
requires
$m=1\>, N_{ij}=\epsilon_{ij}$ or $N_{ij}= n^2\delta_{ij}$ if the
medium is isotropic.

Once the latter identifications have been made, everything  proceeds
as  in  the case of a massive particle in  a  $3+1$  dimensional
curved space-time. The momenta are given by

\be p_i = \frac\hbar\lambda m\frac{N_{ij}\dot x^j}
{\sqrt{N_{ij}\dot x^i\dot x^j}}
\label{13}\ee
The  invariance  under reparametrization of the  action  (\ref{12})
gives rise to a constraint:

\be p^2=\frac{\hbar^2}{\lambda^2}m^2
\Leftrightarrow N^{ij}p_ip_j - \frac{\hbar^2}{\lambda^2}m^2=0
\label{14}\ee

It is interesting to point out that the constraint above completely describes
 the classical dynamics. In fact, following Landau
\cite{[Landauclasica]} we can replace $p_i$ in (\ref{14}) by
$\frac{\p S}{\p x^i}$ and obtain a Hamilton-Jacobi-like equation

\be N^{ij}\p_i S\p_j S - \frac{\hbar^2}{\lambda^2}m^2=0 \>,\label{15}\ee
for which the general solution contains all the information about the classical
dynamics of the system \cite{[Goldstein]}.

Obviously, equation
(\ref{15}) leads to the same dynamics irrespective of the
interpretation  one gives to the classical
theory. In the next section we shall see that this is
no longer the case in the quantum theory. Different interpretations of
the {\it same} classical equations lead to different quantum
theories.  In the present case only the interpretation in 1)
leads to the correct quantum theory.

\section{Quantization.}

In this section we  quantize
Geometric Optics in two different ways resorting to a massive
Klein-Gordon field and a Proca field, respectively
[the Dirac field  is not considered here;
the interested reader can find the relevant expressions in Ref.
\cite{[Birrell]}].
The two different interpretations of the classical
theory discussed above lead, as we shall see, to different quantum theories. We
show that only one of these can be interpreted as the correct stationary
Maxwell theory. As stated above we will deal with both
interpretations at the same time by considering the more general case of
a particle with site-dependent mass and moving in a
curved space(-time).
In this and next section, indices are raised and lowered
with the metric $N_{ij}$.

\subsection{Quantization as a scalar field.}

As is well known, to quantize the system in (\ref{12})
{\it \`a la} Klein-Gordon
one introduces a complex scalar field $\phi$ and makes use,
in eq. (\ref{14}),
 of the
basic quantization rules: $p_i\rightarrow -i\nabla_i$
 to obtain for $\phi$ the equation of motion

\be -\hbar^2\square\phi -
\frac{\hbar^2}{\lambda^2}m^2\phi - \hbar^2\alpha R\phi=0\>\label{16}\ee
or

\be  [\square +\frac{m^2}{\lambda^2}+\alpha R]\phi=0\label{17}\>.\ee
Here $\alpha$ is a, in principle unknown, dimensionless constant;
R is the scalar curvature, and
$\square$  the Laplacian operator associated to the
metric $N_{ij}$,

\be \square \equiv \nabla^i\nabla_i=
\frac1{\sqrt{N}}\p_i\left(N^{ij}\sqrt{N}\p_j\right)\label{18.a},\ee
$N$ being the determinant of the metric.

The equation of motion (\ref{17}) can be obtained
from the action

\be S_{KG} =
\int\d^3x\sqrt{N}\left[-N^{ij}\p_i\phi\p_j\phi^*+
\frac{m^2}{\lambda^2}\phi\phi^*+\alpha R\phi\phi^*\right]\>.\label{18}\ee

Note that Plank's constant has disappeared from the quantum
equation of motion (\ref{17}) for $\phi$. Unlike the
mechanical case, quantizing Geometric Optics does not require the
introduction of the Plank constant but another parameter $\lambda$ which
apparently plays a quite different role.  (In the following, we
shall drop the parameter $\hbar$ from the ``mass"
$\frac\hbar\lambda m$).

\subsection{Quantization as a Proca field.}

The action of a Proca field $A^i$, with mass $\frac{m}\lambda$,
in a curved space with metric $N_{ij}$ is given by
{\cite{[Birrell],[Ryder]}:

\be S_P=\int \d^3x \sqrt{N}\left[
-\frac14F^{ij}F_{ij}+
\frac12\frac{m^2}{\lambda^2}A^iA_i+\frac12\kappa R A^iA_i+
\frac12\gamma R_{ij} A^iA^j\right]\label{19}\ee
where
\be F_{ij}=\p_iA_j-\p_jA_i= \nabla_iA_j-\nabla_jA_i\>,\label{20}\ee
and $R_{ij}$, $R$ are respectively the Ricci tensor and the scalar curvature
associated to the metric $N_{ij}$. As in the scalar case, the dimensionless
constants $\kappa$ and $\gamma$ which appear
in these terms are, in principle, undetermined.
The equations of motion are easily obtained with the result

\be \p_i\sqrt{N}F^{ij}+\sqrt{N}(\frac{m^2}{\lambda^2}\delta^j{_k} +
\kappa R\delta^j{_k} +\gamma R^j{_k})A^k=0\>\label{21}\ee
or, what is equivalent,

\be \nabla_iF^{ij}+(\frac{m^2}{\lambda^2}\delta^j{_k} +
\kappa R\delta^j{_k} +\gamma R^j{_k})A^k =0\label{22}\ee
Eq. (\ref{22}) implies the following equations of
motion for the basic field $A^i$

\be \square A^j -\nabla^j(\nabla_k A^k) - R{^j}{_k}A{^k} +
(\frac{m^2}{\lambda^2}\delta^j{_k} +
\kappa R\delta^j{_k} +\gamma R^j{_k})A^k =0\>.
\label{24}\ee

Equation (\ref{21}) implies in addition

\be \p_j(\sqrt{N}(\frac{m^2}{\lambda^2}\delta^j{_k} +
\kappa R\delta^j{_k} +\gamma R^j{_k})A^k) = 0\Leftrightarrow
\nabla_j((\frac{m^2}{\lambda^2}\delta^j{_k} +
\kappa R\delta^j{_k} +\gamma R^j{_k})A^k)=0\label{23}\ee

\section{Physical interpretation.}

In this section we shall compare the quantum theories constructed
above with the Maxwell theory. We will see that only the
first interpretation of the classical theory gives the
correct quantum one, which coincides with the
stationary Maxwell theory in a material medium. On the contrary, the
interpretation of light rays as  massive particles moving in a
curved space does not lead to the correct wave theory.

Maxwell equations in material media without sources read
\cite{[Landaucontinuos]}:

\bea \rot\> \E = -\frac1c\frac{\p \B}{\p t}\>,
&&\qquad  \div\> \B =0\>,\label{25}\\
 \div\> \D = 0, &&\qquad \rot\> \H = \frac1c\frac{\p \D}{\p
t}\>,\label{26}\eea
which must be completed with the {\it constitutive relations}

\be D^i=\sum_j\epsilon_{ij}E^j\>,
\qquad B^i=\sum_j\mu_{ij}H^j\>.\label{26.a}\ee

In most material media the relevant constants are $\epsilon_{ij}$
since $\mu_{ij}\approx 1$. This is the only case
that we will consider in this paper.

The stationary equations are obtained by replacing
$\frac{\p}{c\p t}$ everywhere with $\frac{-i}\lambda$, and are given by:

\be \frac{i}{\lambda}\B = \rot\> \E  \>,\qquad
\rot\> \H = -\frac{i}\lambda\D \label{stationary}\ee

Let us consider the action

\be S_{Maxwell} =
\int\d^4x\{ -\frac14F_{\mu\nu}F^{\mu\nu}\} \>,\label{35}\ee
where indices are raised and lowered with the metric

\be \d s^2=\frac{c^2}{n^2}\d t^2 - \d {\bf r}^2=
{\bf v}^2\d t^2 - \d {\bf r}^2\>.\label{34}\ee

With the identifications

\be F_{\mu\nu}=\left(\ba{cccc}0&E_x&E_y&E_z\\
-E_x&0&-B_z&B_y\\-E_y&B_z&0&-B_x\\-E_z&-B_y&B_z&0\ea\right),
F^{\mu\nu}=\left(\ba{cccc}0&-D_x&-D_y&-D_z\\
D_x&0&-H_z&H_y\\D_y&H_z&0&-H_x\\D_z&-H_y&H_z&0\ea\right)\>,
\label{35.b}\ee
we get the correct Maxwell equations (\ref{25},\ref{26})
(and constitutive relations (\ref{26.a}))
in an isotropic medium with $\epsilon=n^2$ and $\mu=1$.

The action in (\ref{35}) differs from Maxwell action in a curved
space-time in the volume element only.
If we restrict the action in (\ref{35}) to the stationary case,
$\frac{\p^2}{c^2\p t^2}\equiv-\frac{1}{\lambda^2}$, and fix the gauge
by putting $A_0=0$, we obtain the
action in (\ref{19}) with a flat metric
$N_{ij} = \delta_{ij}$, and  a ``mass"
$m=\frac n\lambda$.
We see then that the first interpretation of the classical theory of light,
when quantized, leads to the correct quantum theory
which
is the stationary Maxwell theory in a material medium.

The identities in (\ref{35.b}) imply the identifications

\be D^i=-\frac{i}{\lambda} n^2A_i\>,\label{27}\ee
\be E^i=-\frac{i}{\lambda}A_i\label{28}\>,\ee
\be B^i=\frac12\sum_{jk}\epsilon^{ijk}F_{jk}=
(\rot\> {\bf A})^i\>,\ee
\be H^i=\frac12\sum_{jk}\epsilon^{ijk}F^{jk}\>,\ee
which give, of course, the stationary Maxwell equations in an
isotropic medium (\ref{stationary}),
together with the expected
constitutive relations ($\mu=1$),

\be \D=n^2\E\>,\qquad\B=\H\>.\ee

Another proof of the equivalence above can
 be obtained by eliminating in  (\ref{stationary})
the magnetic field ${\bf B}$. We are led this way  to a equation of motion
for ${\bf E}\propto {\bf A}$, that coincides  with eq. (\ref{24})
with a flat metric
and mass $m=n/\lambda$:

\be \nabla^2 E^i -\p_i(\p_j E^j) + \frac{n^2}{\lambda^2}E^i=0\>.\ee

\subsection{Is Quantum Geometric Optics a theory of fields in curved space?}

Let us now
consider the second interpretation of Geometric Optics.
Here $m=1$ and $N_{ij}$ is
expected to be related to the dielectric tensor $\epsilon_{ij}$
which, for isotropic media, is given by
$\epsilon_{ij}=n^2\delta_{ij}$.

Let consider first the scalar case. The flat-metric interpretation gives for
the scalar field $\varphi$ the equation of motion

\be  -\nabla^2\varphi -\frac{n^2}{\lambda^2}\varphi=0\label{Hemholtz}\>,\ee
equation which is known in the optical literature as Hemholtz's equation. The
curved-metric interpretation leads, in place, to the wave equation

\be -\square\phi -\frac1{\lambda^2}\phi -\alpha R\phi=0 \label{Hemholtz2}\>.\ee

Let us restrict our atention
to isotropic media,  which is
the only case that eq. (\ref{Hemholtz}) can take into account. The Ricci tensor
and the scalar curvature for a metric of the form
\be
\d s^2=n^2\delta_{ij}\d x^i\d x^j\>\label{conformalmetric}\ee
are given  by

\be R_{ij}= -\frac1n\p_i\p_jn + 2\frac{\p_in\p_jn}{n^2}-
\delta_{ij}\frac{\p_k\p_kn}{n}\>,\label{Riccitensor}\ee
and
\be R = \frac1{n^2}\left\{-4\frac{\p_i\p_in}{n} +
2\frac{\p_in\p_in}{n^2}\right\}
\label{scalarcurvature}\>.\ee
It it easy to check that if we make in (\ref{Hemholtz}) a change of field
$\varphi=\sqrt{n}\phi$ we obtain for $\phi$
the equation (\ref{Hemholtz2}) with the specific
value $-\frac18$ for  $\alpha$. Hence, for the scalar-field case,
both interpretation
of the classical theory lead to the same quantum theory. The interpretation
with curved metric is more general since it can also take
into account non-isotropic media.
The value of $\alpha$ is, of course, the one which corresponds,
for massless scalar
fields,  to a conformally invariant coupling with the metric \cite{[Birrell]}.
This way of proceeding
can serve as a guide to deal with the Proca field.

The flat-metric equations of motion for the Proca field are (in the following
repeated indices are summed over except indicated otherwise):

\begin{eqnarray} \p_iF_{ij} +
\frac{n^2}{\lambda^2}A_i=0\quad;\quad
\p_i(n^2A_i)=0\label{Procaflat}\>.\end{eqnarray}

In the isotropic case, the equations of motion for the curved-space Proca field
are ($G_{ij}=\p_iC_j-\p_jC_i$):

\bea 0&=&\p_i G_{ij} - \frac{\p_in}{n}G_{ij}+
\frac{n^2}{\lambda^2}C_j\nonumber\\
&&+\kappa\left\{-4\frac1n\p_i\p_in + 2\frac{\p_in\p_in}{n^2}\right\}C_j
\label{Procaisotropic}\\
&&+\gamma\left\{-\frac1{n}\p_i\p_jn + 2\frac{\p_in\p_jn}{n^2}-
\delta_{ij}\frac{\p_k\p_kn}{n}\right\}C_i\nonumber\>.\eea

Let us make in (\ref{Procaflat}) a change of fields $A_i=n^sC_i$.
We obtain

\bea 0&=&\p_i G_{ij} +s\frac{\p_in}{n}G_{ij} +
\frac{n^2}{\lambda^2}C_j\nonumber\\
&&+s\left\{\frac{\p_in}{n}\p_iC_j-\frac{\p_jn}{n}\p_iC_i\right\}
\label{Procaflat2}\\
&&+s(s-1)\frac{\p_in}{n}\left\{\frac{\p_in}{n}C_j-\frac{\p_jn}{n}C_i\right\}
+s\left\{\frac{\p_i\p_in}{n}C_j-\frac{\p_i\p_jn}{n}C_i\right\}\nonumber\>,\eea
and
\be 0=\p_iC_i +(s+2)\frac{\p_in}{n}C_i\>.\ee
It is easy to check that no choice of $\kappa$, $\gamma$ and $s$
 can make the
wave equations  (\ref{Procaisotropic}) and (\ref{Procaflat2}) to coincide.

In conclusion, we can state that Quantum Geometric Optics {\it is
not} a theory of fields in curved space.

We should point out here that the conclusion above
is not in contradiction with the
result obtained by Plebanski \cite{[Plebanski]}. He showed that
 Maxwell theory in a curved space-time can be identified with
the same
theory in a flat space-time but evolving in a material medium.
The latter identification
 requires, nevertheless,
constitutive relations
 that are different from the ones we have been dealing with
in this paper:
 $D^i =\epsilon_{ij}E^j$  and ${\bf H}={\bf B}$.

\section{The optical analogy of Classical Mechanics.}

As is well known \cite{[Goldstein]}, the trajectories with fixed
energy $E$ for
a time-independent mechanical system with kinetic energy
 $T=\frac12M_{ij}(q){\dot q^i}{\dot q}^j$ and potential $V$
can be obtained from the action

\be S = 2\int \sqrt{E-V}\sqrt{T}\d \tau\>.\label{36}\ee
The parameter $\tau$  has
no physical relevance, since the action in (\ref{36}) is
reparame- trization invariant. This principle, which we shall refer to
as the Maupertuis principle, is the mechanical analogue of the Fermat
principle. Here $\sqrt{E-V}$ plays the role of the refraction index.

Let us consider a mechanical system for which
$M_{ij}=m\delta_{ij}$ with $m$ constant (the mass of the
particle). We shall see that  canonical
quantization of the action  in (\ref{36}) gives the right
time-independent Schr\"odinger equation.

The canonical momenta are given by

\be p_i=\frac{\sqrt{(E-V)}}{\sqrt{T}}{m{\dot q}^i}\>.\label{37}\ee
The constraint associated with the reparametrization invariance of the
action is

\be \sum_i p_ip_i=2m(E-V)\>.\label{38}\ee

If we apply the basic quantization rules
$p_i\rightarrow -i\hbar \frac{\p}{\p x^i}$  to eq. (\ref{38})
we find the stationary Schr\"odinger equation

\be -\hbar^2\nabla^2\Psi -2m(E-V)\Psi=0\>.\label{39}\ee

We see then that the exact analogue of the time-independent
Schr\"odinger equation for
Optics is eq. (\ref{17}), the Hemholtz equation,
which can be written in another way:

\be  -\lambda^2\nabla^2\phi -n^2\phi=0\label{200}\>.\ee
Since $n^2$ is always positive, the solutions of eq. (\ref{200})
 belong to the continuous spectrum, that is to
say, they are scattering solutions
and hence
not normalizable.

There is  a simple procedure to go from the usual
Hamilton Principle to the
Maupertuis Principle and vice versa. Let us  start with the
action in the form that is required by the Hamilton Principle:

\be S= \int \d t\left\{\frac12M_{ij}{\dot q^i}{\dot q^j}-(V-
E)\right\}\>,\label{100}\ee
where we have added a convenient total derivative
$\frac{d}{d t}Et$ to the usual Lagrangian.
Let us now introduce an arbitrary parameter $\tau$ to describe
the trajectories of the system. We have $t=t(\tau),\> \d t = t'\d
\tau$ (prime indicates derivative with respect to the parameter
$\tau$). The
action (\ref{100}) takes the form

\be S=
\int \d \tau\left\{\frac12\frac{M_{ij}{q'^i}{ q'^j}}{
t'}-{t'}(V-E)\right\}\>.\ee
In order to obtain a reparametrization-invariant system we now hide
the relationship between $\tau$ and $t$ by replacing $t'$ with a new
quantity, the {\it vielbein} or {\it einvein} $e$. We obtain

\be S= \int \d \tau\left\{\frac12\frac{M_{ij}{q'^i}{ q'^j}}e
-e(V-E)\right\}\>.\label{101}\ee
The physical meaning of $e$ is that of
volume (length), or metric, along the trajectories:
$\d t^2 = e^2 \d \tau^2$.

The action in (\ref{101}) is, in fact, equivalent to the action that appears in
the
formulation of the Maupertuis principle. The quantity $e$  plays the
role of a Lagrangian multiplier, and can be eliminated by using
its equation of motion:

\be 0 = \frac{\delta L}{\delta e}
= -\frac{T}{e^2} -(V-E)\Rightarrow e=\sqrt{\frac{T}{E-V}}
\label{101.b}\ee
If we replace in (\ref{101}) $e$ by its value in
(\ref{101.b}) we obtain the action in (\ref{36}).

\section{The case of true reparametrization-invariant systems.}

The procedure scketched in the previous section to obtain a
reparametrization-invariant system from a mechanical one can
be applied the other way round. However, the procedure when
applied in this  direction,
is not well
defined: it does not lead to a unique result.
The reason for this ambiguity
is that the same reparametrization-invariant action
can be obtained from different mechanical ones. This problem can be
traced back to the  fact that in the reparametrization-invariant
action (\ref{36}) there is no way of distinguishing between
kinetic and potential energy.

In order to ilustrate these points,
let us consider, very briefly, 2-dimensional
induced gravity which is
a true reparametrization-invariant system.
The action, after being restricted to the spatially homogeneous
minisuperspace, reads \cite{[Wavefunctions]}:

\be S= \int\,\d \tau\left[\frac{1}{2}\frac{a}{e}\Phi'^2 +
2\frac{{a'}\Phi'}{e} +
\frac{\Lambda}{2}e a\right]\>.\label{105}\ee
Nevertheless, as stated above, this action is not unique:
the same dynamics can
be obtained from other actions related to (\ref{105})
by a redefinition of $e$.
If we get rid of the Lagrange multiplier $e$ we obtain

\be S= 2\int\d\tau \sqrt{\frac\Lambda2 a}
\sqrt{\frac{1}{2}a\Phi'^2 +
2{a'}{\Phi'}}\>.\label{106}\ee

The Hamiltonian constraint obtained from (\ref{105}) can be written
as\footnote{Note that in the
literature of Quantum Cosmology what we call quantum Hamiltonian
constraint is named Wheeler-DeWitt equation.}

\be -\frac{1}{8}ap_a^2 + \frac{1}{2}p_ap_\Phi -
\frac{\Lambda}{2}a=0\>. \label{107}\ee
However, since there is no  preferred choice of $e$ here, the
Hamiltonian constraint can be written in different forms,
for instance,

\be -\frac{1}{8}(ap_a)^2 + \frac{1}{2}ap_ap_\Phi -
\frac{\Lambda}{2}a^2=0\>. \label{108}\ee
In fact, there are physical reasons which make this
latter form of the constraint preferable \cite{[Wavefunctions],[Teitelboim]}.

For Geometric Optics there are also
a series of actions involving a Lagrange multiplier ({\it einvein})
$e$ which are equivalent to the Fermat action (\ref{1}).
Here, as it happens for mechanical systems,
there is a prefered choice of parametrization, $\tau=x^0$,
begin $x^0$ the physical time,  which obeys

\be  {\bf \dot x}^2 = \frac1{n^2}\>.\ee

This preferred choice of parametrization singularizes, among all possible
actions, the one for which the
 interval $\d s^2=e^2\d \tau^2$ has the meaning of physical time:
 \be \d s^2\equiv\d (x^0)^2\>.\ee
This preferred action is

\be S=\frac{\hbar}{\lambda}\int\d\tau\left\{n^2\frac{{\bf{\dot x}}^2}{e} +
e\right\}\>.\ee
However, the existence of
this singularized action
does not help us since it
does not distinguish between the two interpretation of the classical theory
studied above. In fact, it point to
the wrong direction since it seems to
indicate that the good interpretation is the curved-metric one.

To summarize, we can say that the existence of a true
time enables us to distinguish kinetic energy
from potential energy in the classical action.
This fact provides a preferred form of the
{\it classical} Hamiltonian constraint which, in general, makes
the ordering problem of its quantum counterpart, the Schr\"odinger
equation, less poisonous, even though the problem is not completely solved.

\section{Final remarks.}

The structure of the present paper have been somewhat circular:
we began by
considering Geometric Optics from the point of view of Mechanics
and found the analogy fruitful (Sections 2, 3 and 4). Then,
in Sec. 5,
we considered Mechanics from the point of view of Geometric
Optics and again found the analogy fruitful.
As
a result, we have studied in depth the classical and quantum
dynamics of Geometric Optics along with their relationship with Mechanics.
We have widely shown that there is a close analogy between
Optics and Mechanics and that between them, at a certain level, a complete
isomorphism can be established.
This explains the success of  mechanical
techniques when applied to Optics \cite{[Wolf],[Sternberg]}.

The present paper ilustrates both the power and the weakness of
the quantum theory in its present form.  For instance,
we can obtain the stationary
Maxwell theory from the Fermat principle. However, the Fermat
principle, together with the canonical quantization
procedure in its present form, does not indicate clearly
which is the correct quantum
theory. The same situation appears in connection with the Maupertuis
principle and the time-independent Schr\"odinger equation.
We need some information, a great amount of
information indeed, that is provided neither
by the classical theory nor the
quantization procedure.
These examples
clearly ilustrate the difficulties which appear when quantizing true
reparametrization-invariant systems such as gravity theories \cite{[Isham]}.
In fact, the quantization procedure provides neither the correct
equations of the quantum theory nor the physical interpretation
we should give to it if we were able somehow to obtain
these equations.
\v 1cm
\noindent{\bf Acknowledgements.} M. Navarro is grateful to the
Spanish MEC for a postdoctoral FPU grant.
J. Guerrero thanks the Spanish MEC for a FPU grant. M. Navarro is
grateful to J. Navarro-Salas for very useful comments and suggestions.

\end{document}